\documentclass[conference, letter]{IEEEtran}
\IEEEoverridecommandlockouts

\ifCLASSINFOpdf
  \usepackage[pdftex]{graphicx}
  \graphicspath{img}
  \DeclareGraphicsExtensions{.pdf,.jpeg,.png}
\else
\fi

\usepackage{cite}
\usepackage{mathtools}
\usepackage{lineno}
\usepackage{amsmath,amssymb,amsfonts}

\usepackage{graphicx}
\usepackage{textcomp}
\usepackage{algorithm} 
\usepackage{algpseudocode} 
\usepackage{tikz}
\usetikzlibrary{angles, quotes, arrows.meta}
\usepackage{siunitx}
\usepackage{longtable,tabularx}
\usepackage[left=1.29cm, right=1.29cm, top=1.9cm, bottom=4.29cm]{geometry}
\usepackage{enumitem}
\usepackage{anyfontsize}
\usepackage{multicol}
\usepackage{multirow}
\usepackage{etoolbox}
\usepackage{fancyhdr}
\makeatletter
\renewcommand{\fnum@figure}{Figure \thefigure}
\makeatother

\title{ {A Simulation-Optimization Framework for Developing Wind-Resilient AAM Networks} \\

\thanks{This research was funded by Supernal, grant number 052838-002.}
}

\author{\IEEEauthorblockN{Emin Burak Onat$^*$, Shangqing Cao, Raiyan Rizwan, \\Xuan Jiang, Mark Hansen, Raja Sengupta}
\IEEEauthorblockA{University of California, Berkeley \\
Berkeley, CA, USA \\
$^*$corresponding: eminburak\_onat@berkeley.edu}
\and
\IEEEauthorblockN{Anjan Chakrabarty}
\IEEEauthorblockA{Supernal \\
Fremont, CA, USA}  
}

\renewcommand\IEEEkeywordsname{Keywords}
\IEEEaftertitletext{\vspace{-1\baselineskip}}

\begin{document}

\maketitle
\thispagestyle{fancy}

\noindent \begin{abstract}
Environmental factors pose a significant challenge to the operational efficiency and safety of advanced air mobility (AAM) networks. This paper presents a simulation-optimization framework that dynamically integrates wind variability into AAM operations. We employ a nonlinear charging model within a multi-vertiport environment to optimize fleet size and scheduling. Our framework assesses the impact of wind on operational parameters, providing strategies to enhance the resilience of AAM ecosystems. The results demonstrate that wind conditions exert significant influence on fleet size even for short-distance flights, their impact on fleet size and energy requirements becomes more pronounced over longer distances. Efficient management of fleet size and charging policies, particularly for long-distance networks, is needed to accommodate the variability of wind conditions effectively.
\end{abstract}

\vspace{0.3cm}


\section{Introduction}

The emergence of AAM has changed urban transportation, with electric vertical take-off and landing (eVTOL) aircraft promising efficient, sustainable mobility~\cite{b1}. However, the operational efficiency of these systems is highly restricted by environmental factors, notably wind conditions, which pose significant challenges to flight scheduling and overall network performance \cite{b2, b3, b4}. Existing models do not adequately account for wind dynamics, or they neglect to integrate a comprehensive, multidimensional charging strategy that includes critical variables such as timing, location, the volume of charging tasks, and charging intervals. These elements are essential for the practicality and functionality of AAM systems \cite{b5, b6}. This oversight can lead to inaccuracies in estimating network capacity, safety thresholds, and the overall resilience of the system. Such limitations highlight the pressing need for an inclusive framework capable of precisely simulating and optimizing AAM operations amidst the added variability of wind conditions.



This paper introduces a framework aimed at advancing AAM operational planning by integrating charging and scheduling strategies that consider wind variability. Our model dynamically optimizes flight and charging schedules within a multi-vertiport system, and computes performance metrics using an advanced simulation environment with models for infrastructure, batteries, chargers, wind, and eVTOL aircraft dynamics. The primary objective of our work is to explore the critical nexus between optimal fleet management and the impact of wind variability on AAM operations, illuminating not merely the isolated effects of wind on AAM operations but more importantly, how these effects scale and interact with optimal scheduling and charging processes. By simultaneously addressing these components, our study aims to bridge a gap in current research, offering insights into how AAM networks can adapt to and incorporate environmental uncertainties into their operational planning.



\section{Related Work}

AAM network simulation research spans diverse methodologies, from analyzing eVTOL energy constraints to comparing air taxis with traditional taxicabs, and integrating traffic simulations to evaluate urban impacts \cite{b7, b8, b9}. Additionally, research has explored the potential shift in commuter demand towards AAM, factoring in travel time savings and vertiport transfer times \cite{b11}, and employed collaborative system-of-systems modeling to optimize AAM operations covering demand forecasting, vertiport design, and air traffic management \cite{b10, b12}.


Efficiency in AAM operations is dependent on dispatching logic, scheduling strategies, and fleet management. The importance of dispatching logic has been highlighted, pointing to the need for optimized decision-making processes in managing AAM operations \cite{b13}. Researchers have also developed optimal scheduling strategies and approach control models for multicopter VTOL aircraft \cite{b14}. Scheduling has been further addressed by presenting a heuristic approach for arrival sequencing and scheduling for eVTOL aircraft, focusing on the on-demand nature of AAM services \cite{b15}.

The effective implementation of AAM requires addressing the challenges wind variability poses to eVTOL networks. While substantial progress has been made in optimizing infrastructure and scheduling to service eVTOL networks, the scalability of these solutions under diverse wind conditions—ranging from steady winds to severe gusts—has received less attention. These wind conditions can cause longer flight times, increased energy consumption, and necessitate larger battery capacities, impacting fleet management and network flow. Research has highlighted the challenges posed by varying wind conditions, including the establishment of wind category thresholds to guide aircraft separation and takeoff/landing protocols in the absence of specific federal aviation standards \cite{b2}. Analysis based on a decade of historical wind data indicates a marked decrease in performance at vertiports when wind is factored in. Another study focuses on mitigating the effects of wind by applying wind-optimal lateral trajectories in a high-physical-fidelity simulation environment \cite{b16}. The study finds a significant increase in flight duration and energy consumption when flying in a headwind, but an insignificant decrease when flying along a wind-optimal versus great circle path. The paper also highlights the importance of a wind field’s spatial variability and relative route direction in its impact on energy and time. \cite{b2} notably misses the impact of wind direction and thus spatial variability in a multi-vertiport eVTOL network. \cite{b16} focuses on the impact of wind on a single route as opposed to a network handling hundreds of daily flights. 

To the best of our knowledge, there have been no studies on the impact of wind on infrastructure, fleet sizing and scheduling for holistic eVTOL network operations in the context of an environment that simultaneously simulates wind direction, speed, and spatial and temporal resolution. Our research aims to address these gaps by conducting a detailed analysis of how varying wind conditions, based on five years of historical data, impact fleet size and network performance. We focus on optimizing demand scheduling through an integrative framework that encompasses the comprehensive modeling of passenger, aircraft, and energy flows. This work will be an important step in helping stakeholders understand the resources required for eVTOL networks to safely and efficiently tolerate realistic wind conditions year-round. 

\section{Contributions}
In this work, we make three key contributions:
\begin{enumerate}
    \item \textbf{AAM network simulation environment:} A simulation environment that models eVTOL network operations, including environmental factors like wind, to assess their impact on fleet dynamics, energy usage, and network efficiency.
    \item \textbf{Optimization model for dynamic scheduling and charging:} Extended an optimization model \cite{b20} to consider time-varying flight durations and energy needs due to wind.
    \item \textbf{Methodology to integrate optimization libraries with discrete event simulation: } We propose a methodology to integrate offline optimization with discrete-event simulation, facilitating practical application of optimization algorithms for real-world AAM operational planning and resource management.
\end{enumerate}

\section{Network Simulation-Optimization Framework}
\begin{figure}[!htb]
  \centering
  \includegraphics[width=0.45\textwidth]{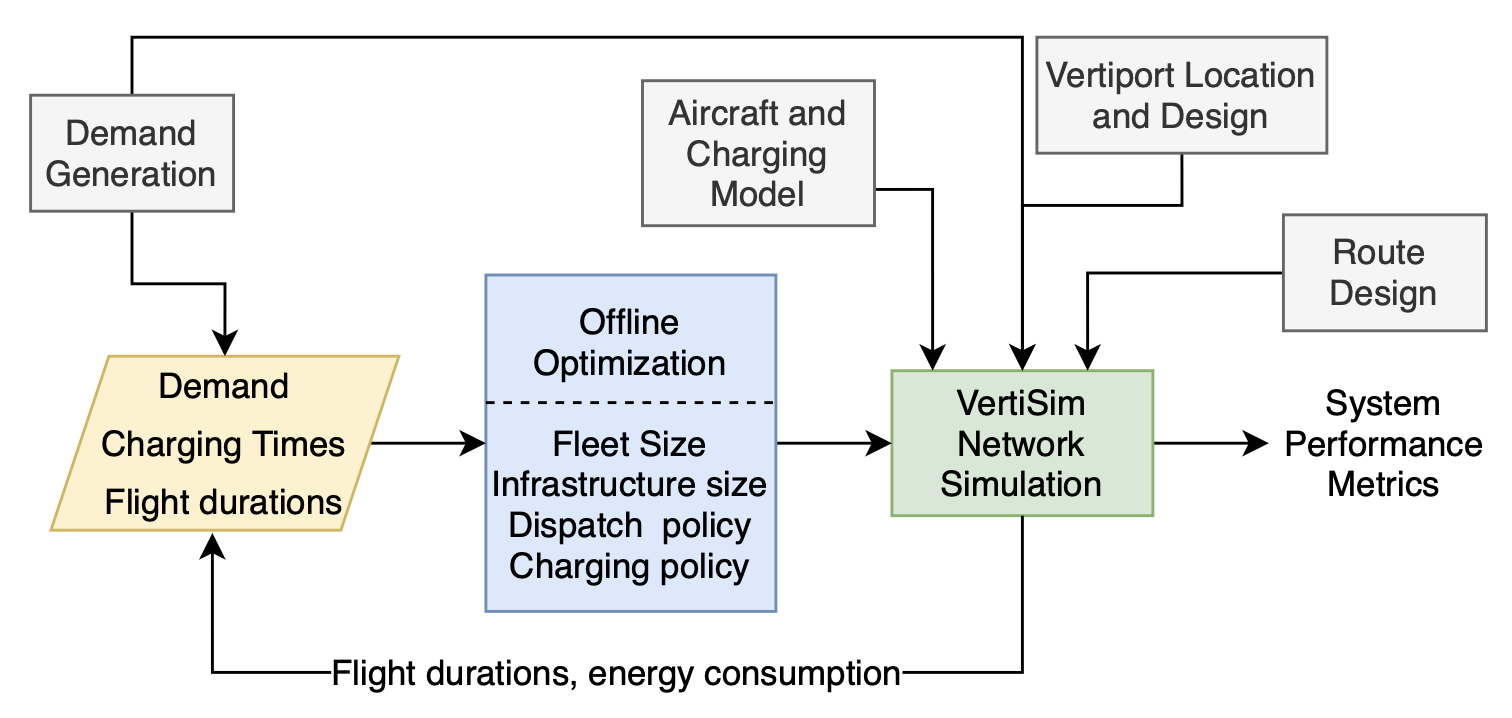}
  \caption{Simulation - Optimization framework}\label{fig:framework}
\end{figure}

This section presents a novel framework that integrates the capabilities of VertiSim, an event-driven simulator for AAM networks, with an offline optimization tool. VertiSim provides a detailed simulation of vertiport operations and aircraft movements within a AAM network, the optimization tool complements it by computing essential operational parameters such as fleet size, infrastructure requirements, flight schedules, and charging policies. The framework's architecture is illustrated in Figure \ref{fig:framework}. Key inputs for the optimization include flight demand, flight times for each route that are pre-computed by the simulator, and a charging timetable delineating the charging duration for various SoC levels. A passenger arrival process is modeled and pre-processed to establish a flight schedule that aims to deliver a specific service level. VertiSim integrates a suite of inputs, such as the aircraft and charging models, flight routes, vertiport layouts and locations, raw passenger arrival process, and pre-optimized flight and charging schedules, to simulate network operations. The resulting performance metrics encompass aircraft utilization rates, average aircraft occupancy, vertiport throughput and resource usage, energy consumption across different flight phases, number of repositioning flights, and passenger waiting and trip times.

\subsection{Agent-Based Vertiport Network Simulator}
Expanding on our foundational VertiSim work on AAM network simulation \cite{b17}, this paper introduces VertiSim's integration with optimization libraries for improved decision-making and a wind module to simulate environmental effects on AAM operations.

\begin{figure}[!htb]
  \centering
  \includegraphics[width=0.49\textwidth]{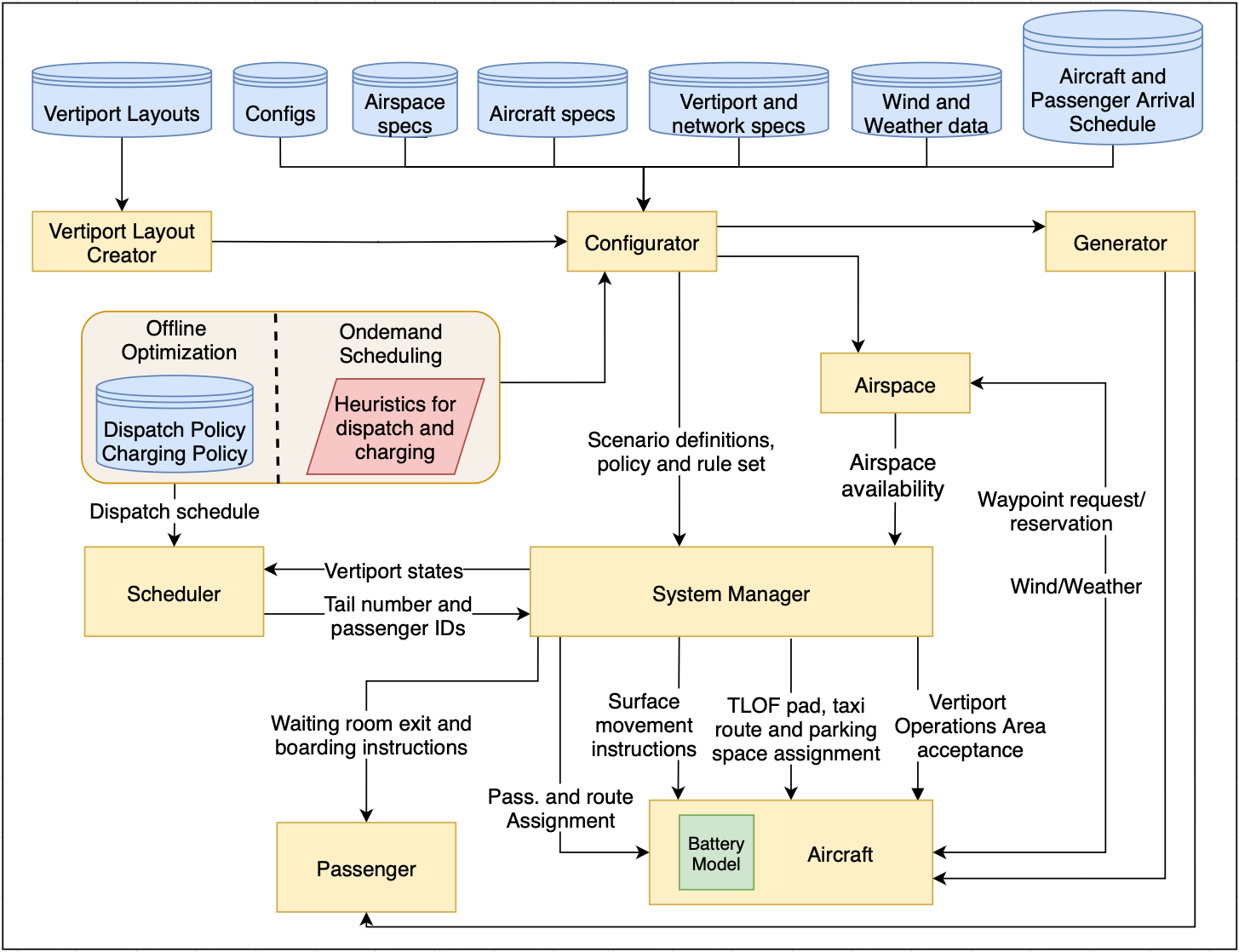}
  \caption{VertiSim software architecture}\label{fig:vertisim_arch}
\end{figure}

The new software architecture of VertiSim is illustrated in fig. \ref{fig:vertisim_arch}. The \textit{Vertiport Layout Creator} generates a node-link graph modeling vertiport structures—TLOF areas, parking pads, chargers, and taxiways. The \textit{Configurator} refines these entities, assigning server roles for queues and setting rates for chargers and security checkpoints. \textit{Generator} creates aircraft and passenger agents based on the simulation clock and inputs from the configurator, setting up their initial state, origin, destination, and passenger arrival times. The \textit{Airspace} module creates a 3D navigable airspace node-link graph for eVTOLs.

An \textit{Aircraft} within the system is characterized by a set of state variables that include its location, horizontal and vertical speeds, state of charge (SoC), identifiers for its departure and arrival vertiports, the duration of service at its current position, the list of passengers on board, its current operational state and its assigned priority level. Passengers appear at vertiports post-check-in, wait for the \textit{Scheduler} to assign trips, and follow the \textit{System Manager} to gates, exiting the simulation post-disembarkation at the destination vertiport.

The \textit{Scheduler} module is tasked with monitoring the count of \textit{Passengers} in each waiting room, initiating trips based on pre-defined criteria. A trip is initiated when either the quantity of waiting passengers matches the aircraft's capacity, or an individual passenger's wait time exceeds a specified limit. Continuing from previous studies, VertiSim continues to support on-demand services, which are directed by heuristics for dispatching and charging. The integration of an optimization library has led to significant modifications in the functionality and structure of both the Scheduler and the System Manager modules. The Scheduler has adopted a new role, implementing the dispatch and charging strategies as determined by the optimization results.

The \textit{System Manager} commands the simulation's resource allocation and routing, ensuring aircraft-trip assignments in line with the \textit{Scheduler}'s directives.

\subsubsection{Wind Model}
VertiSim models seven flight phases: hover climb, climb transition, climb, cruise, descent transition, descent, and hover descent. The aircraft's vertical velocity is maxed out in every phase except cruise, where it is zero. This imposes a required time of arrival (RTA) on the aircraft's horizontal position to ensure that it reaches its target (latitude, longitude) and target altitude at the same time step. Thus, the aircraft's necessary horizontal ground speed is fixed for every phase except cruise. VertiSim employs true airspeed for power calculations, and ground speed to determine flight time. Energy consumption is a combination of both, since it is the integral of power over time.

\paragraph{\textbf{Cruise Phase}}
We focus on operating at minimum power during cruise since we assume there is no RTA. The eVTOL maintains its maximum range airspeed while still flying in the right direction with respect to the ground. This is essentially crabbing at a power optimal speed. The diagram below describes how we accomplish this; in principle, we are able to change the direction of the aircraft without changing its airspeed, such that the wind pushes it toward its destination vertiport.

\begin{figure}[!htb]
  \centering
  \includegraphics[width=0.35\textwidth]{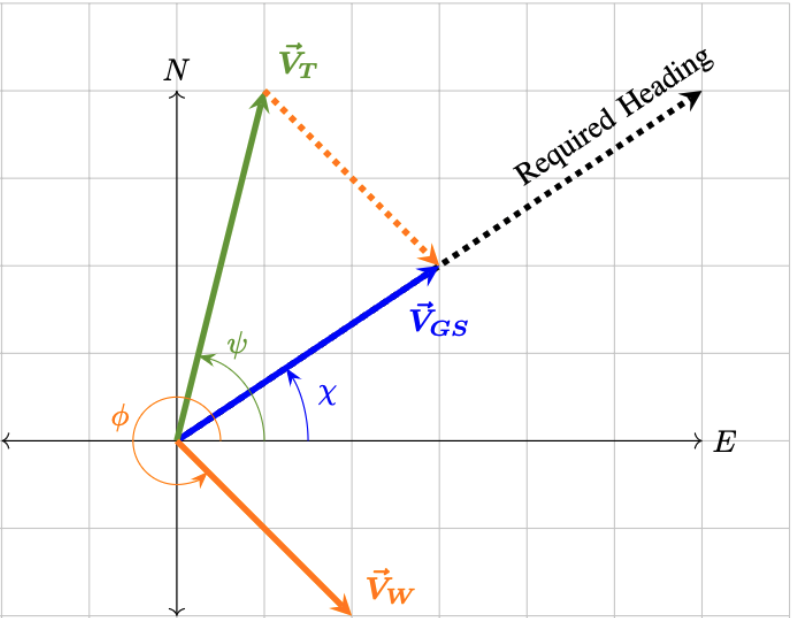}
  \caption{A visual representation that illustrates the interaction between the aircraft's true air velocity ($\Vec{V}_T$), ground velocity ($\Vec{V}_{GS}$), wind velocity ($\Vec{V}_{W}$), angle of $\Vec{V}_W$ ($\phi$) and the strategic adjustment of the aircraft's heading ($\psi$) to maintain course towards the destination vertiport in the presence of wind.}\label{fig:aircraft_heading}
\end{figure}

\vspace{-5mm}
\begin{equation}\label{eq:wind_angles}
\sin(\chi - \psi) = \frac{|\Vec{V}_W|}{V_C} \sin(\phi - \chi)
\end{equation}

The relationship between the required heading, airspeed, and wind velocity is captured by Equation~\ref{eq:wind_angles}. By solving this equation analytically, we calculate $\psi$. This allows us to define the aircraft's true velocity vector as $\langle V_{C}\cos(\psi), V_{C}\sin(\psi) \rangle$, aligning with the power-efficient cruise speed, $V_C$. Consequently, the eVTOL is steered towards its destination vertiport in the presence of wind $\Vec{V}_{W}$. VertiSim has checks to make sure our calculations are accurate within floating point errors: 

\begin{center}
$\angle{\Vec{V}_{GS}} = \chi$ \hspace{1cm} $|\Vec{V}_T| = V_C$  \hspace{1cm} $|\Vec{V}_{GS}| >= V_{threshold}$
\end{center}
The first equality states that the angle of the ground speed vector ($\Vec{V}_{GS}$) has to match the intended heading towards the destination vertiport. The second checks whether the aircraft maintains a constant airspeed equal to its optimal cruise speed. The third inequality is a soft check; if violated, the calculation from the other phases is applied to cruise. 

For cruise, the true airspeed remains the same as that with zero wind. Thus, the power consumption is also the same. However, the aircraft will fly slower or faster with respect to the ground depending on the angle of the crosswind. This means that the energy consumption will still change. While the RTA approach could be applied to cruise (decreasing flight duration at the expense of power consumption), we found this technique to consume less energy overall.

\paragraph{\textbf{Other Phases}}
In all flight phases other than cruise, the aircraft maintains a constant ground speed by adjusting its power output to counteract the effects of wind. VertiSim does not account for vertical wind effects, requiring adherence only to the formula: $V_h = D_h / (D_v / V_v)$ where \textit{V} indicates velocity, \textit{D} distance, \textit{v} vertical, and \textit{h} horizontal. 

The ground velocity is subject to the constraint that $\Vec{V_{GS}} = \Vec{V_{T}} + \Vec{V_{W}}$, leading to $\Vec{V_{T}} = \Vec{V_{GS}} - \Vec{V_{W}}$. As a result, while wind conditions do not alter the duration of the eVTOL's flight, they do affect its power needs and, consequently, its energy consumption.

\subsection{Optimization Model}

The optimization model yields a flight schedule and a charging policy that minimizes the number of aircraft required to satisfy a given flight demand profile. We consider a finite set of discretized time intervals at which control actions can be taken. Table \ref{tab:notations} introduces the notations used in the optimization model. We formulate the objective function as the following:
\begin{align}
\min &\sum_{i}\sum_{k} n_{i}^{k}(t=0) + \sum_{i} \sum_{x} \sum_{y} C_i^{xy}(t = 0) \nonumber \\
     &+ \alpha \cdot  \sum_{t} \sum_{i} \sum_{j} \sum_{k} u_{ij}^{k}(t) 
\end{align}

The first two summation terms in the objective function represent the total number of eVTOL aircraft, based on time step $t=0$, i.e., at the beginning of an operating day. Because no aircraft is flying at $t=0$, the sum of aircraft idling and charging at $t=0$ constitutes the fleet size. The third term in the objective function represents the total number of flights with a small weight $\alpha$, which has the purpose of constraining the solution space so that between two solutions that yield the same fleet size, the solution that minimizes the number of flights is preferred. We set $\alpha = 0.00001$ because $\alpha$ has to be small enough such that $\alpha \cdot \textit{number of flight}$ does not exceed 1.

\begin{table}
    \centering
    \caption{Optimization Model Notations}
    \renewcommand{\arraystretch}{1.5} 
    \begin{tabular}{c|p{6.6cm}}
     \textbf{Notation} & \parbox{5cm}{\centering \textbf{Explanation}}   \\ \hline 
     $C_{i}^{xy}(t)$ &  No. of aircraft at vertiport \(i\) that begin to charge at time step \(t\) with an initial SoC of \(x\) and a target SoC of \(y\)  \\ \cline{1-2}
     \(u_{ij}^{k}(t)\) &  Number of aircraft departing for vertiport \(j\) from vertiport \(i\) at time step \(t\) with SoC \(k\)  \\ \hline
     
     \(n_{i}^{k}(t)\) &  No. of idle aircraft at vertiport \(i\) of SoC \(k\) at time step \(t\)  \\ \cline{1-2}
     \(f_{ij}^{t}\)  &  No. of flights departing from vertiport $i$ to vertiport $j$ at time step $t$ required to satisfy travel demand  \\ \cline{1-2}
     \(\gamma_{k}\) & Charging time needed to transition from \(SoC_{k-1}\) to \(SoC_{k}\)  \\ \cline{1-2}
     \(K\) & Number of SoCs after discretization  \\ \cline{1-2}
     \(\tau_{ij}^{t}\) & Flight time from \(i\) to \(j\) at time step $t$ in time steps  \\ \cline{1-2}
     \(\kappa_{ij}^{t}\) & Reduction in SoC for a flight from vertiport \(i\) to \(j\) departing at time step $t$  \\ \cline{1-2}
     $T$ & Planning horizon $T = \text{number of time steps} + \max\limits_{ijt} \tau_{ij}(t) + 1$ \\ \cline{1-2}

     $\mathcal{V}$ & Set of vertiports  \\ \cline{1-2}

     $\mathcal{A}_{ij}^{t}$ & $\{t^\prime \in \{1, \dots, T\} |  t^\prime+\tau_{ij}(t^\prime) = t\}$ \\
     
    \end{tabular}
    \label{tab:notations}
\end{table}

Minimization of the objective function is subject to the following constraints:
\begin{align}
    &n_{i}^{k}(t) =  n_{i}^{k}(t-1) + \sum_{j\in\mathcal{V}-\{i\}} \sum_{t^\prime \in \mathcal{A}_{ji}^{t}} u_{ji}^{k+\kappa_{ji}^{t^\prime}}(t^\prime) - \sum_{j\in\mathcal{V}-\{i\}} u_{ij}^{k} (t) \notag\\
    & + \sum_{x=0}^{k-1} C_{i}^{xk} (t-\sum_{i=x+1}^{k} \gamma_{i}) - \sum_{y=k+1}^{K} C_{i}^{ky}(t), \quad \forall i,t 
     \label{cons:dynamic_equation}
\end{align}
\begin{align}
    &\sum_{k \in\{1, \cdots, K\}} u_{ij}^k(t) \geq f_{ij}^{t}, \quad \forall i, j, t 
    \label{cons:demand}
\end{align}
\begin{align}
    u_{ij}^0(t) = 0, \quad \forall i,j,t 
   \label{cons:energy}
\end{align}
\begin{align}
    &n_{i}^{k}(0)=n_{i}^{k}(T), \quad u_{ij}^{k}(0)=u_{ij}^{k}(T), \quad C_{i}^{xy}(0)=C_{i}^{xy}(T) \quad  \nonumber \\  & \forall i,j,k,x,y
    \label{cons:sta}
\end{align}

\vspace{0.3em}

The dynamic equation that governs the evolution of the system is given by constraint \ref{cons:dynamic_equation}, which models the state of the aircraft in relation to the charging policy and the flight schedule. We model the number of aircraft that enter a state of idle at each vertiport at each time step of a certain SoC as a combination of the following terms: (A) the number of idling aircraft that are carried over from the previous time step, (B) the number of aircraft that will arrive at vertiport $i$ at time step $t$ of SoC $k$, (C) the number of aircraft that depart from vertiport $i$ of SoC $k$, (D) the number of aircraft that complete charging to SoC $k$ at time step $t$ at vertiport $i$, and (E) the number of aircraft that are committed to charging at vertiport $i$ starting from SoC $k$. Note that we parameterize the flight duration and energy consumption with respect to time in the model, an advancement compared to the previous model \cite{b20}. Such design enables us to model the time-varying impact that wind has on the system within the optimization time horizon, reflecting the realistic impact of wind patterns during an operating day. Constraint \ref{cons:demand} ensures that flight demand for each vertiport pair $(i,j)$, measured as the required number of flights $f_{ij}^{t}$, is satisfied at each time step. Constraint \ref{cons:energy} ensures that aircraft cannot fly when the SoC level equals 0, which is the reserve SoC. Constraint     \ref{cons:sta} ensures that the number of aircraft in different states and SoCs are the same at the beginning and the end of the time horizon. 

\subsection{Simulation-Optimization Integration}
The optimizer generates flight schedules with location and State of Charge (SoC) for each flight, while charge schedules include location, current SoC, and target SoC without specific aircraft identifiers. The \textit{System Manager} manages aircraft assignments for trips and charging by matching available aircraft to these criteria. To conserve computational resources, the optimizer functions on a 5-minute decision-making interval, which leads to discrepancies with the simulator's higher fidelity and event-driven nature. This discrepancies arises from differences in the discretized 5-minute intervals used by the optimizer for flight and charging durations compared to the varying durations in the simulator influenced by factors like queueing delays and aircraft weight.

\begin{figure}[!htb]
  \centering
  \includegraphics[width=0.45\textwidth]{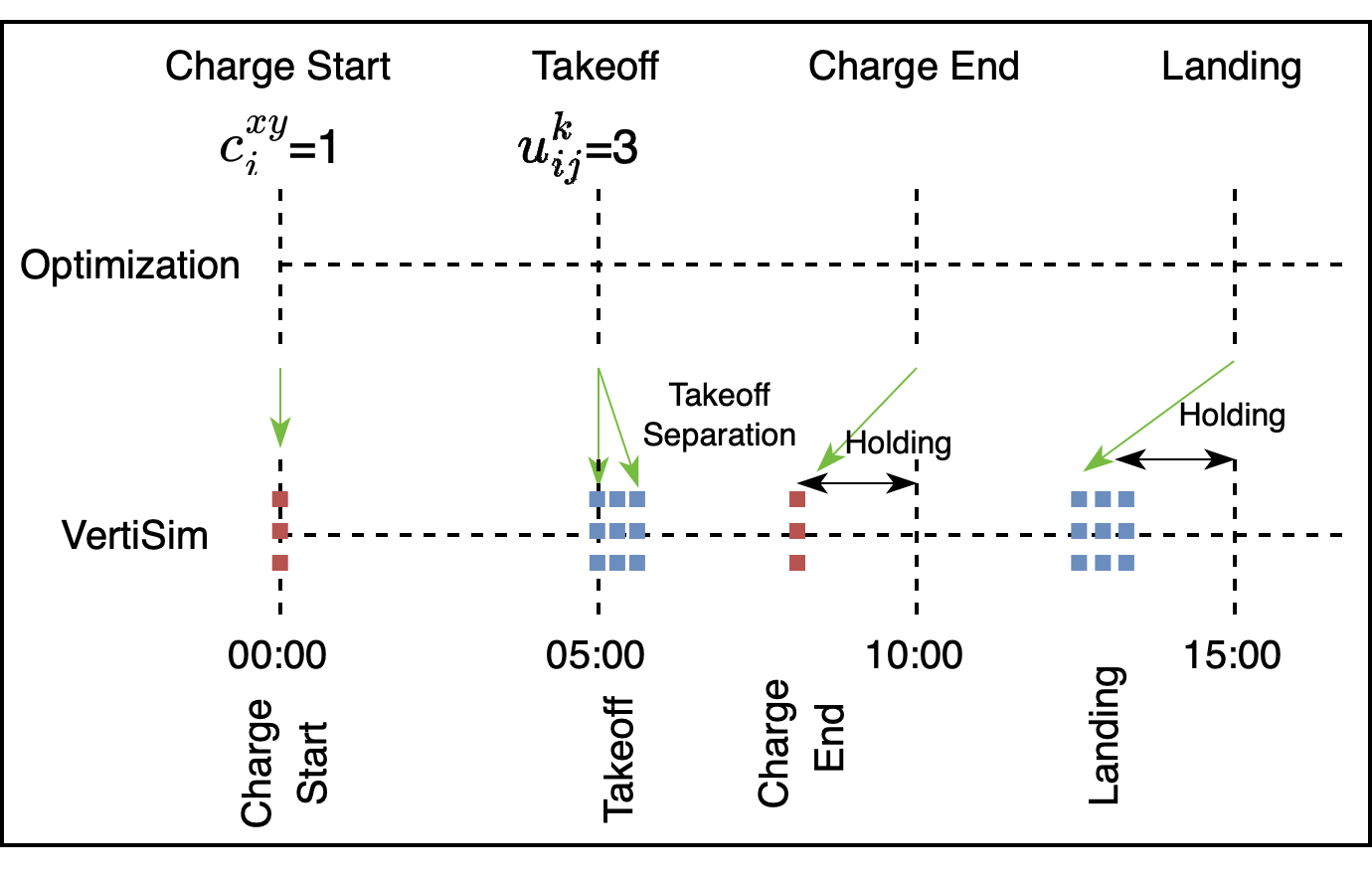}
  \caption{Simulation - Optimization integration illustration example}\label{fig:integration}
\end{figure}

An example scenario is illustrated in fig. \ref{fig:integration}. The optimizer schedules charging to start at 00:00, and 3 takeoffs are set at 05:00, with the charging ending at 10:00 followed by the end of flights at 15:00. However, in VertiSim processes like charging, takeoff, and landing are not fixed to 5-minute blocks and can occur at any simulated time. Additionally, VertiSim internally applies separation protocols to aircraft. Consequently, with a single FATO layout selected for the vertiport, it cannot accommodate two operations (takeoff or landing) simultaneously. To reconcile these temporal discrepancies, VertiSim implements a holding strategy: if an aircraft finishes charging or a flight segment earlier than the optimization schedule dictates, it enters a idling pattern until the scheduled time arrives. This is indicated by the blue dotted lines that represent the actual event times in VertiSim, which precede the solid black lines of the scheduled times in the optimization framework. This strategy allows VertiSim to  pause and synchronize its activities with the optimization schedule without altering the sequence or integrity of events.

\section{Analysis Design}
\subsection{Vertiport Networks and Area of Study}
This paper explores the impact of wind on operational metrics across a two-vertiport network. The choice of this network is driven by its simplicity, which allows for a focused analysis free from the complexities of larger networks. The area of study spans 150 miles from Monterey to Sacramento, employing hypothetical vertiport distances, ranging from 20 to 150 miles in 10-mile increments. In this context, Vertiport A is identified as the Monterey vertiport. Vertiport B, on the other hand, is a variable designation that shifts to reflect the changing distances, representing the counterpart vertiports at each specified distance increment. The heading angle from Vertiport A to Vertiport B is calculated to be 13 degrees relative to north, indicating a flight direction approximately from south-southwest to north-northeast.

\subsection{Passenger Demand Modelling}
\begin{figure}[h]
  \centering
  \includegraphics[width=0.5\textwidth]{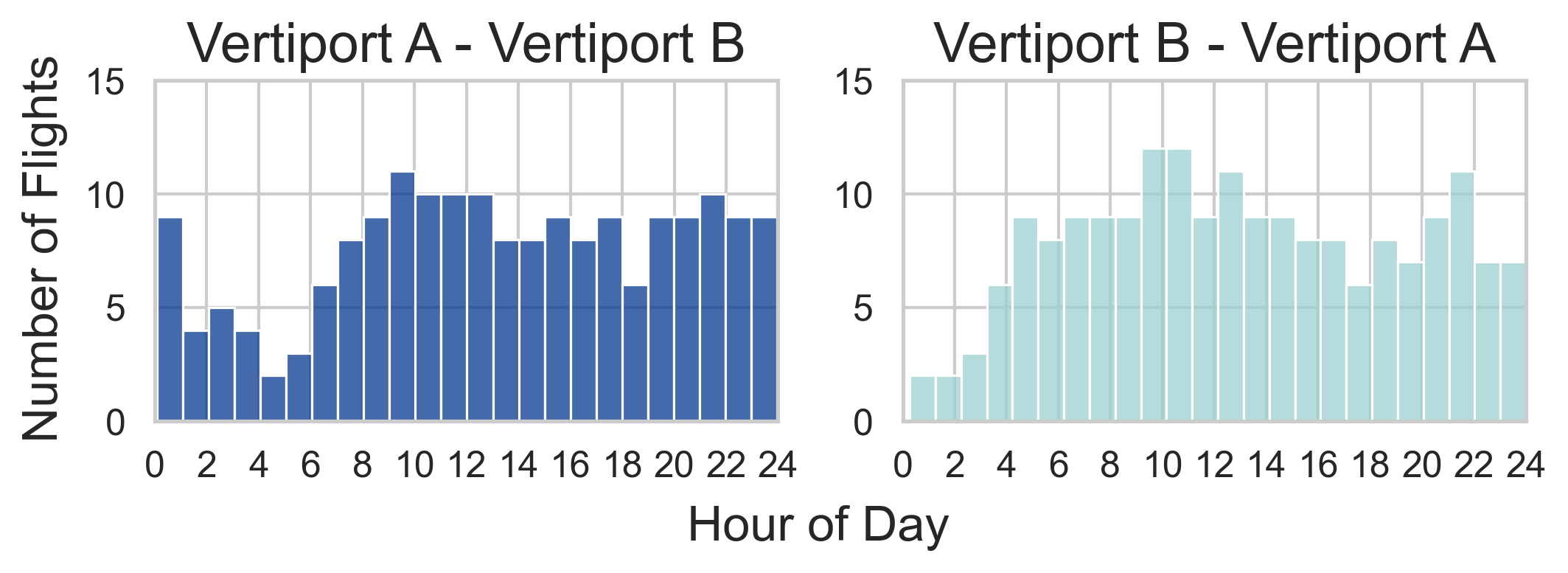}
  \caption{Distribution of flight demand for the analysis}\label{fig:demand}
\end{figure}

In this two-vertiport system, AAM demand is derived using the process outlined in \cite{b20}. We assume the airport vertiport is Vertiport A and the other vertiport is Vertiport B. We set the autoregressive coefficient $\alpha$ in the demand generation model to 0, thus not considering the variation in demand. We set the expected average directional demand ($ADD$) to 500 and generate a realization of passenger arrivals for 1/1/2019. To obtain demand in number of flights, we assume that a flight is created when there are four passengers waiting at the vertiport or if the first passenger has waited 5 minutes. Eventually, we obtain a flight demand for each flight direction. Fig. \ref{fig:demand} shows the flight demand in the two directions of flight.

\vspace{-1mm}

\subsection{Charging and Aircraft Model}
We employed the same charging and aircraft model as outlined in \cite{b17} for our simulations. The charging model was developed using experimental data from electric vehicle to represent eVTOL charging processes, given the lack of specific eVTOL charging information. The modeled charger maintains a constant charge power up to a 20\% SoC, after which the charge power decreases linearly until the battery is fully charged. To compute the cumulative energy delivered over time, we integrated the charging power with respect to time. This total energy consistent with the aircraft's battery capacity, determined to be 160 kWh as detailed in \cite{b17}. This model underscores the inherently non-linear characteristics of the charging process, where charging speed decreases as the battery approaches full capacity.

The power requirements for takeoff and landing (eq. \ref{eq:takeoff_landing}), climb and descent (eq. \ref{eq:climb_descent}), and cruise (eq. \ref{eq:cruise_energy}) are calculated for the selected aircraft model as follows:

\begin{equation}\label{eq:takeoff_landing}
P_{fixed-wing} = [\frac{f W}{FoM} \sqrt{\frac{f W / A}{2 \rho}} + \frac{W V_{climb_v}}{2}]/ \eta_{hover}
\end{equation}

\begin{equation}\label{eq:climb_descent}
P_{fixed-wing} = [W V_v + \frac{1}{2}\rho V^3 S C_{D_0} + \frac{K W^2}{\frac{1}{2} \rho V S}] / (\eta_{climb})
\end{equation}

\begin{equation}\label{eq:cruise_energy}
P_{fixed-wing} = [W V_v + \frac{W V}{[L/D]}] / (\eta_{cruise})
\end{equation}

Here, $K$ is the lift-induced drag coefficient: $K = 1/(4C_{D_0}(L/D)_{max}^2)$. The parameters are defined in table \ref{tab:parameters}.

The aircraft aims for minimum power use during climb, climb transition, descent, and descent transition, while the cruise phase targets maximum range with a desired forward (true air) speed of about 150 mph. For detailed computational aspects please refer to \cite{b18}.

In simulations, passenger weight adjusts the total aircraft weight based on seat occupancy. The lift-to-drag ratio ($L/D$) varies per flight segment – 18 for cruising and 15.6 for climbing and descending. Air density ($\rho$) changes with atmospheric conditions and altitude, assuming standard conditions at ground level.


\begin{table}[!h]
\caption{Aircraft model parameters \cite{b18}}
\centering
\label{tab:parameters}
\small 
\setlength{\tabcolsep}{4pt} 
\renewcommand{\arraystretch}{1.2} 
\begin{tabular}{|c|c||c|c|}
\hline
\textbf{Parameter} & \textbf{Value} & \textbf{Parameter} & \textbf{Value} \\ \hline
MTOM, W & 2182 [kg] & $C_L$ & 1.5 \\ \hline
f & 1.03 & $(L/D)_{max}$ & 15.3 - 18 \\ \hline
$C_{D0}$ & 100 [kg] & $\eta_{hover}$ & 0.85 \\ \hline
Wing area, S & 13 [$m^2$] & $\eta_{climb}$ & 0.85 \\ \hline
FoM & 0.8 & $\eta_{descent}$ & 0.85 \\ \hline
Passenger weight & 0.015 & $\eta_{cruise}$ & 0.90 \\ \hline
\end{tabular}
\end{table}
\noindent\parbox{\linewidth}{
\small
\vspace{1mm}
\textbf{Note:} MTOM: Maximum Take-Off Mass, f: Correction factor for interference from the fuselage, FoM: Figure of Merit, $C_{D0}$: Zero lift drag coefficient, $(L/D)_{max}$: Maximum lift to drag ratio, $\eta$: Efficiency.
}

\subsection{Flight Profile}
We strategically placed airspace waypoints 1 mile apart to ensure a consistent separation among cruising aircraft. These nodes link the initial cruise level of the departure vertiport to the destination's holding unit, forming a direct flight path. Hover fixes are positioned 50 feet above the FATO. Aircraft are modeled to complete their hover climb in 10 seconds, climb transition in 60 seconds at a vertical speed of 300 ft/min and an 6.1\% climb slope, with the climb phase itself lasting 120 seconds at a 550 ft/min rate and a 5.4\% slope. Descent phases, including the descent transition, are set to 120 and 60 seconds respectively, with descent rates of 620 ft/min and 300 ft/min, and descent slopes of 5.4\% and 6.1\% respectively.

\subsection{Wind Data}
The Rapid Refresh (RAP), developed by the National Center for Environmental Prediction, provides hourly weather forecasts with an 8-mile spatial resolution. Utilizing RAP, we extracted five years of historical wind data (2015-2019) at 1,500 feet AGL, using grid points spaced 8 miles apart. We chose the closest grid points along the vertiport route to capture relevant wind data. The wind data shows an average wind speed of 25.5 mph with a predominant direction of 226 degrees relative to the north, indicating winds primarily coming from the southwest. The variability in wind speed and direction is highlighted by a standard deviation of 15.5 mph and 98 degrees, respectively, reflecting significant fluctuations over time.

\subsection{Processing Wind Data for Optimizer Input}


Past research on wind variability and AAM picked specific wind speeds and/or directions as representative of trends in historical data. We take a different approach, in which we abstract away the specific wind directions and speeds by picking clusters that depict distinct snippets of 24-hour, representative energy consumption and flight duration data. This data is generated en-masse from simulating flights in VertiSim across the network in the given wind conditions (pulled from five years of historic data), for every flight direction and distance. As aircraft transition between wind zones, our simulations dynamically adjust to changes in wind speed and direction, thereby ensuring the preservation of wind's spatial and temporal variability in our energy consumption and flight duration calculations without anchoring to specific wind values.

\begin{figure}[t]
\centering
\includegraphics[width=0.5\textwidth]{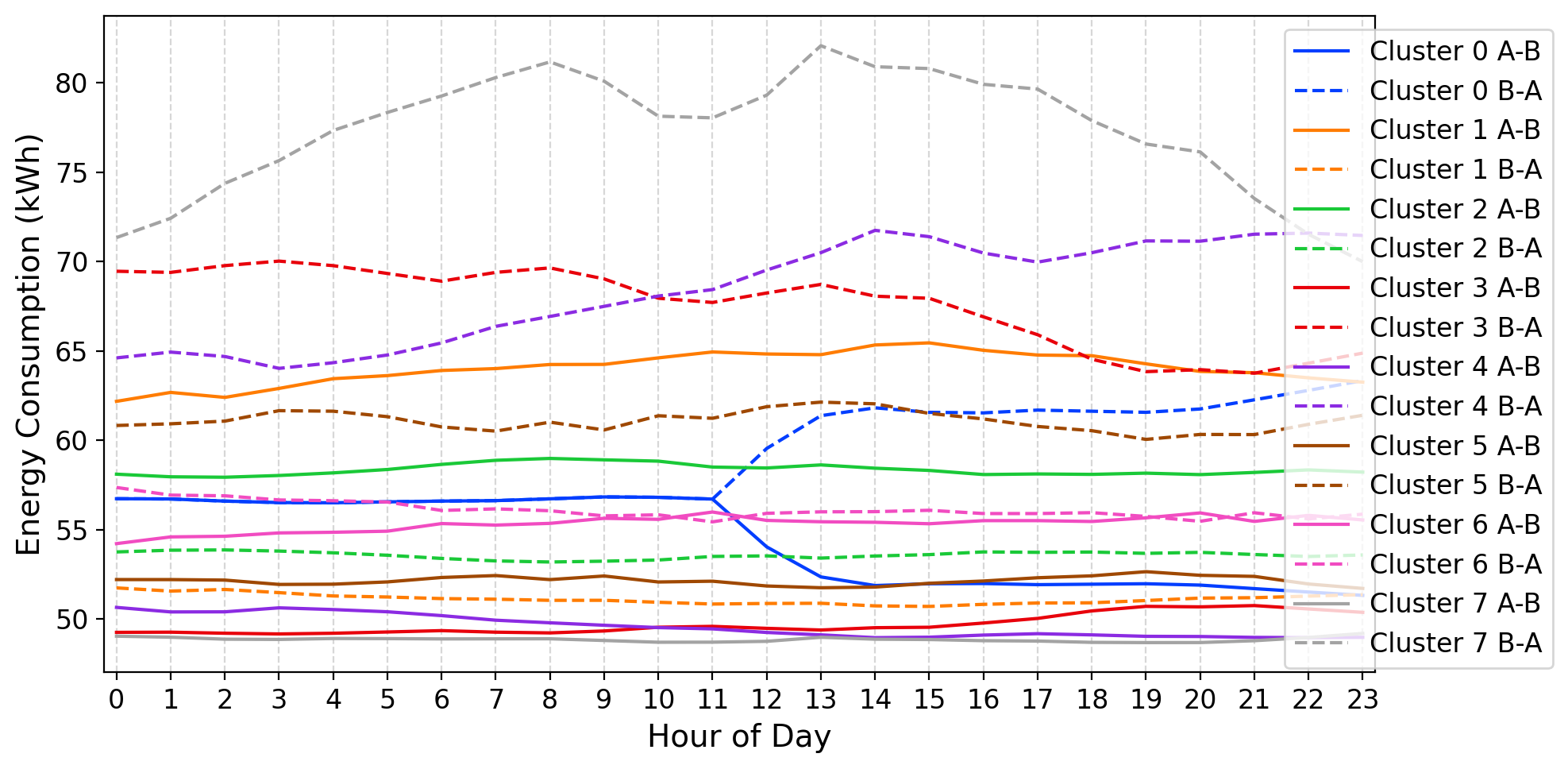}
\caption{Energy consumption clusters for 80 mile distance network. A-B and B-A represent opposite flight directions}
\label{fig:energy_cons_cluster}
\end{figure}

This clustering simplifies the representation of our dataset. We target an explained variation of at least 70\% to determine the optimal number of clusters. Ultimately, we selected the centroids of 8 clusters for each flight direction, with each cluster centroid encompassing a 24-set of flight duration and energy consumption data, representing the varying conditions throughout a day. To exemplify, figure \ref{fig:energy_cons_cluster} demonstrates the energy consumption clusters for an 80-mile network. Similar to energy consumption, flight duration clusters, ranging from 25 to 43 minutes, also exhibit variations that correlate with the changing wind conditions for the same network. Across the entire range of flight distances, the shortest and longest flight durations within our clusters were recorded at 9 and 66 minutes, respectively.

\section{Results}
\subsection{Wind and Flight Distance Impact on Energy Consumption and Charging Policy}
\begin{figure}[t]
\centering
\includegraphics[width=0.48\textwidth]{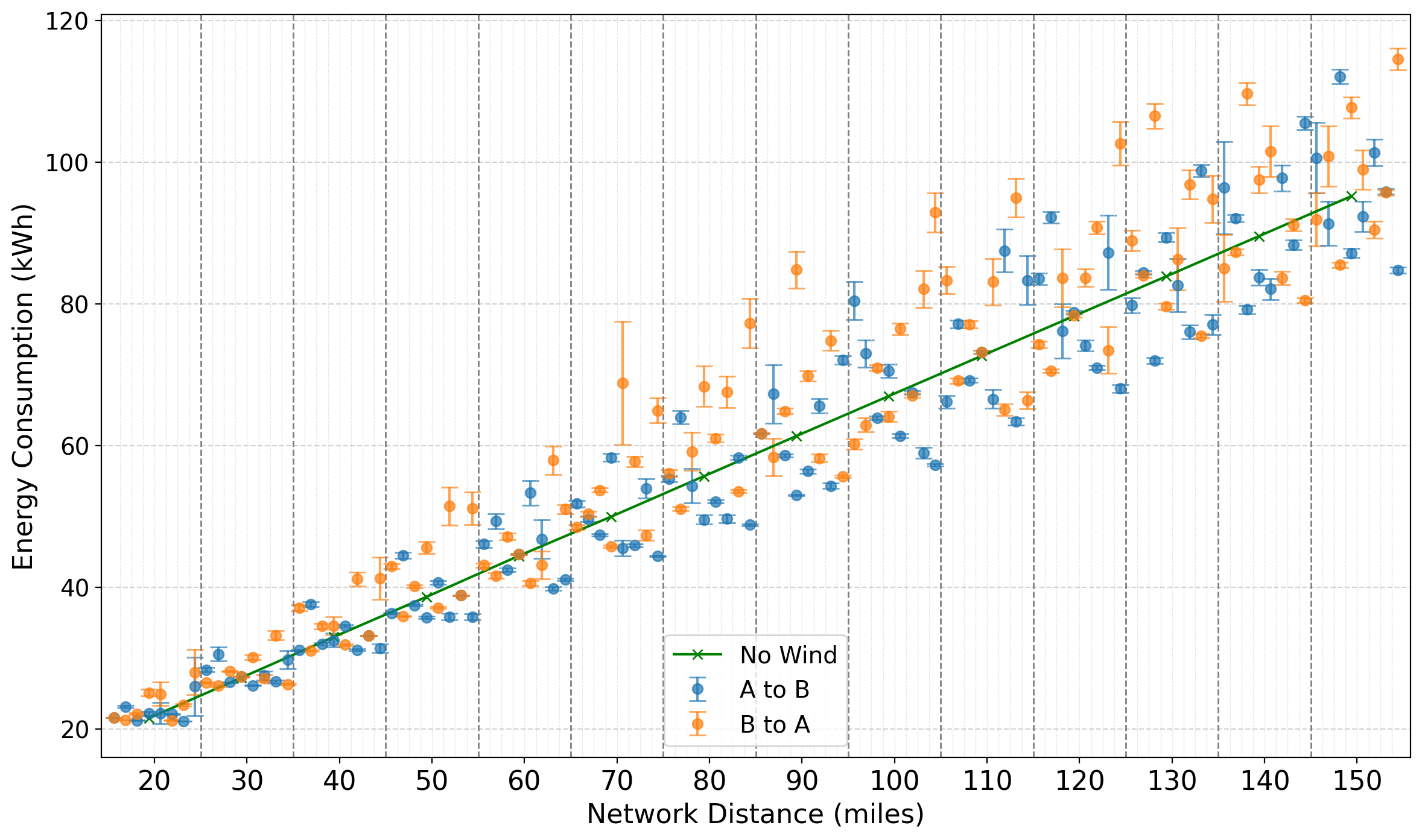}
\caption{Mean energy consumption by cluster within each flight distance, with points indicating mean values and bars showing standard deviation. Data points for different flight directions are plotted along a shared x-axis.}
\label{fig:energy_cons_for_clusters}
\end{figure}

Our study reveals significant differences in energy consumption for varying flight directions and distances, influenced by wind conditions. As shown in Fig. \ref{fig:energy_cons_for_clusters}, energy needs for flights from A to B differ considerably from those for B to A, especially as distance increases. Headwind conditions notably increase energy consumption, requiring up to 43\% more energy compared to no wind conditions. Such increases necessitate more frequent and longer charging sessions which impacts aircraft availability and charging infrastructure requirements. These observations underscore the importance of integrating wind variability into the charging and scheduling model.

\begin{figure}[t]
\centering
\includegraphics[width=0.4\textwidth]{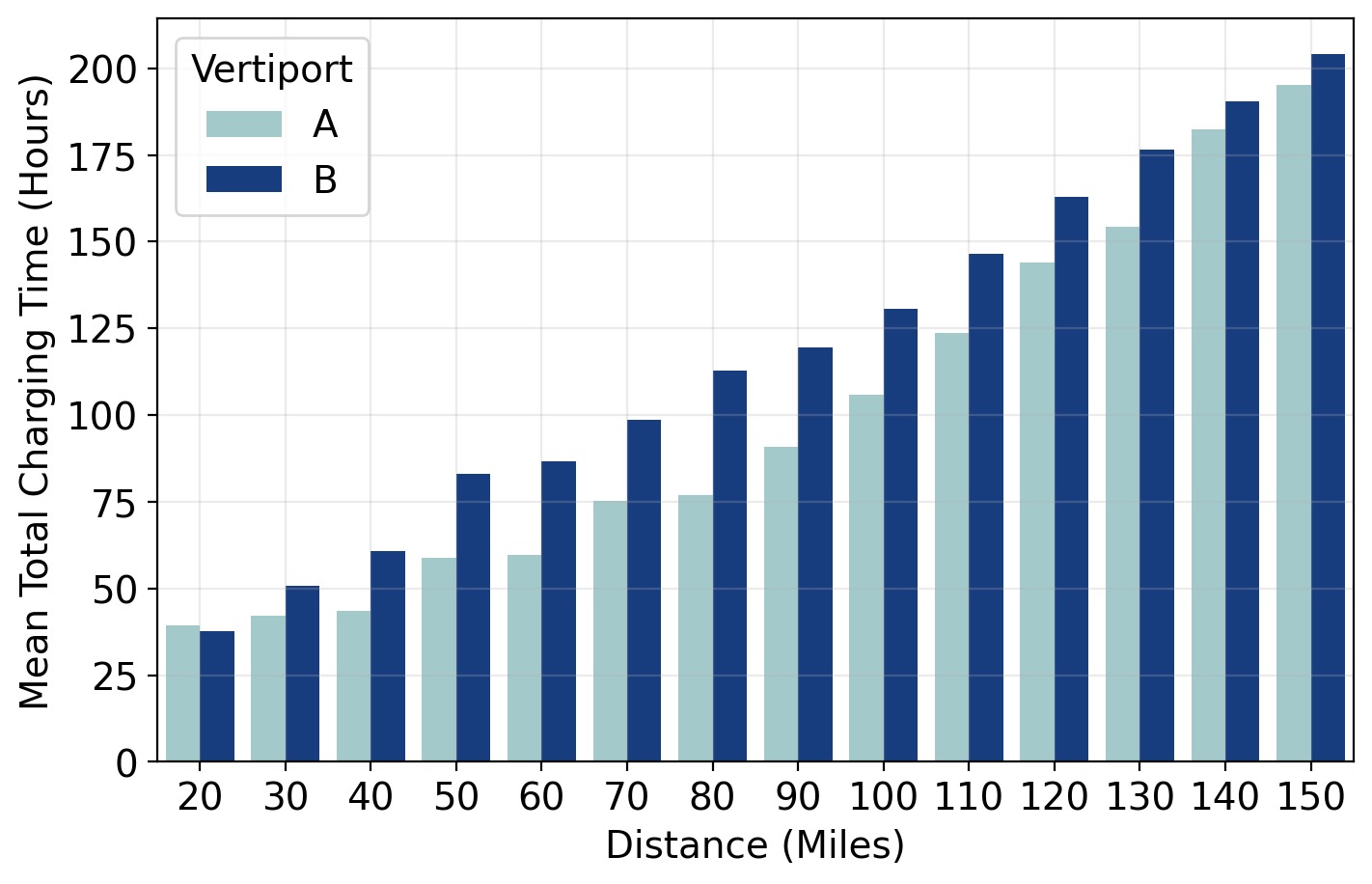}
\caption{Comparison of average total charging times across energy consumption clusters for each flight distance at Vertiport A and B.}
\label{fig:charging_times}
\end{figure}

Considering a 160 kWh battery capacity and the 20\% SoC reserve requirement (32 kWh), it is observed that for distances up to 70 miles, aircraft can usually complete a round trip on a single charge if it starts the trip with a full charge. Beyond this threshold, additional charging becomes a necessity for the return trip. This insight is pivotal for our charging strategy shown in fig. \ref{fig:charging_times}. The mean charging time differential between Vertiport A and B is influenced by the prevalent tailwinds when traveling from A to B. It's observed that the disparity in charging times at each vertiport is minimal for shorter and longer flights. This can be explained by the fact that shorter trips may not necessitate additional charging at either vertiport, while for longer trips, charging is required regardless of the vertiport due to battery depletion. The difference becomes more significant for mid-range distances. This is where the impact of wind conditions on energy consumption is most evident, as it can determine whether a round trip can be completed without the need for recharging at one vertiport, while at the other, charging is indispensable. Understanding this pattern is vital for optimizing the operational efficiency and scheduling of aircraft between these two vertiports.

\subsection{Wind Impact on Fleet Size}
\begin{figure}[t]
\centering
\includegraphics[width=0.5\textwidth]{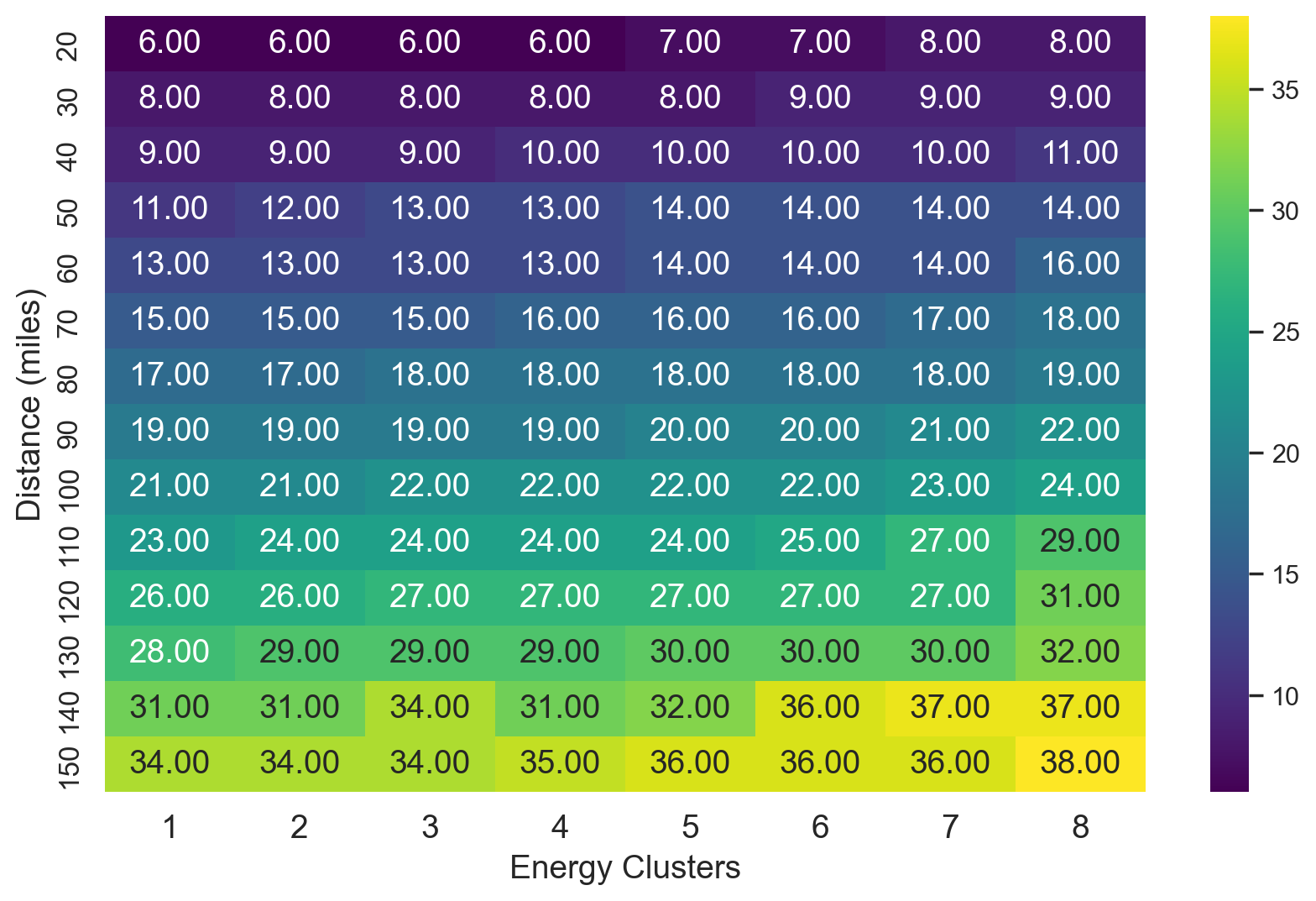}
\caption{Fleet size variability across wind clusters and flight distances. Color-bar represents fleet size.}
\label{fig:heatmap}
\end{figure}

The impact of wind on fleet size for eVTOL operations varies significantly across different distances, as illustrated by the heatmap in Fig. \ref{fig:heatmap}. For short distances up to 40 miles, wind impact causes fleet size variations ranging from 12\% to 33\%, with a difference of up to 2 aircraft across different energy clusters. Although the capital investment increase is small, the proportional increase for the short-distance market is significant. For medium distances between 50 to 100 miles, wind variations cause fleet size increases ranging from 14\% to 27\%, with a difference of up to 3 aircraft across different energy clusters.

For long distances over 100 miles, the influence of wind becomes even more pronounced, suggesting a need for a significant augmentation in fleet size, potentially requiring up to 6 extra aircraft. This incremental, yet non-negligible, increase in fleet size highlights the importance of incorporating wind variability into operational planning for long-distance eVTOL services. While this does not suggest a high degree of volatility, it does indicate that a measured degree of flexibility in fleet management is prudent to maintain service reliability and efficiency in the face of wind-induced variations. The analysis shows that operations over short to mid-range distances are sensitive to wind, requiring precise aircraft management to avoid inefficient fleet utilization. As distances increase, the need for fleet flexibility becomes even more critical to accommodate the heightened impact of wind and ensure that all passengers are served without spillage.



\section{Conclusion and Future Work}
This paper has introduced a comprehensive framework addressing the critical influence of wind on the efficiency of AAM operations. Our findings indicate that wind impacts are substantial for shorter eVTOL routes, with significant variations in fleet size, and become increasingly important as flight distances extend. This particularly impacts charging requirements and fleet management. By incorporating wind variability into our simulation-optimization framework, we have demonstrated the potential to optimize fleet size and ensure efficient scheduling and charging strategies.

For future work, we aim to extend the analysis to more complex networks with multiple vertiports. Another avenue for future work includes the integration of reinforcement learning algorithms into the simulation environment to do online decision-making under wind and demand uncertainties. Ultimately, our goal is to facilitate the deployment of AAM as a safe, efficient, and environmentally friendly mode of urban transportation.


\vspace{12pt}

\end{document}